\RequirePackage{iftex}
\ifLuaTeX
  \RequirePackage{luatex85}
\fi
\documentclass[sigconf,nonacm,natbib=false,leqno]{acmart}
\usepackage{tikz}
\usepackage{braket}
\usepackage{mathtools}
\ifLuaTeX
  \usepackage[haranoaji]{luatexja-preset}
\fi
\setcopyright{none}
\RequirePackage[datamodel=acmdatamodel,style=numeric]{biblatex}
\usepackage{preamble}

\acmConference[EGRAPHS '26]{EGRAPHS 2026}{June 15--19, 2026}{Boulder, Colorado, USA}
\title[Optimizing Optimizations, Declaratively]{Optimizing Optimizations, Declaratively: Optimizing the Higher-Order Functions in Mathematical Optimization with egglog}
\author{Hiromi Ishii}
\email{h.ishii@j-ij.com}
\orcid{0000-0002-7752-1782}
\affiliation{%
  \institution{JIJ Inc.}
  \city{Minato-ku}
  \state{Tokyo}
  \country{Japan}
}
\begin{document}

\begin{abstract}
  We present two applications of egglog~\cite{egglog} to mathematical optimization in JijModeling 2~\cite{jijmodeling:2025,jijmodeling:pypi}, a mathematical modeller whose internal representation is based on simply typed $\lambda$-calculus.

  First, we use egglog to improve \LaTeX{} output for mathematical models expressed with higher-order functions.
  Python comprehensions are desugared into stream operations such as \textsf{map}, \textsf{flat\_map}, and \textsf{filter}; emitting these terms directly produces unnatural mathematical notation.
  We reconstruct comprehension syntax by \emph{ensugaring} higher-order terms and use equality saturation with a custom cost model to minimize temporary variable rebindings.

  Second, we use egglog as a declarative engine for \emph{constraint detection}, extending the previous egg-based approach presented at EGRAPHS '25~\cite{Ishii:aa}.
  Egglog's datalog-style rules let us express multi-step detection logic directly, without external Rust orchestration code.
  We encode parametrized constraints using \emph{Henkin-like constants} and propagate side conditions on subterms and indices through egglog facts.
  Finally, we show that the same ensugaring procedure also reduces large domain-set conditions before saturation, turning a problematic detection case from minutes or nontermination into a few seconds.

  Through these topics, we want to provide an example of an industrial application of egglog, demonstrate the trick to propagate the constraints using the ideas from mathematical logic, and show the importance of optimizing \emph{premises} of egglog rules to get practical performance in egglog programs.
\end{abstract}

\maketitle

\section{Introduction}

\begin{figure*}[!b]
  \[
  \begin{array}{rl}
\text{Problem}\colon &\text{TSP (Quadratic)}\\\displaystyle \min &\displaystyle \sum _{t=0}^{\#C-1}{\sum _{i\in C}{\sum _{j\in C}{{d}_{i,j}  {x}_{t,i}  {x}_{\left(t+1\right)\bmod \#C,j}}}}\\\text{s.t.}&\begin{aligned}
\text{one city}&\quad \displaystyle \sum _{i\in C}{{x}_{t,i}}=1\quad \forall t\;\text{s.t.}\;t\in \left\{0,\ldots ,\#C-1\right\}&\qquad&\text{one time}&\quad \displaystyle \sum _{t=0}^{\#C-1}{{x}_{t,i}}=1\quad \forall i\;\text{s.t.}\;i\in C\end{aligned}
\\\text{where}&\text{Decision Variables:}\\&\qquad \begin{alignedat}{2}x&\in \mathop{\mathrm{TotalDict}}\left[\#C\times C;\left\{0, 1\right\}\right]&\quad &\text{$x_{t,i} = 1$ if City $i$ is visited at time $t$}\\\end{alignedat}\\&\text{Placeholders:}\\&\qquad \begin{alignedat}{2}d&\in \mathop{\mathrm{TotalDict}}\left[\mathrm{C}\times \mathrm{C};\mathbb{R}\right]&\quad &\text{distance between cities}\\\end{alignedat}\\&\text{Category Labels:}\\&\qquad \begin{array}{rl}
C&\text{Labels of Cities}\end{array}
\end{array}
  \]
  \caption{Travelling Salesman Problem\label{fig:tsp}}
\end{figure*}

\emph{Mathematical optimization} is used to model planning tasks such as procurement planning, production planning, and scheduling.
\Cref{fig:tsp} shows a quadratic formulation of the Travelling Salesman Problem (TSP) as a typical example.
The standard workflow is to formulate the mathematical expressions in a \emph{modeller} and pass them to a \emph{solver}.
JijModeling is such a modeller, provided as a Python EDSL and implemented in Rust.
It stores mathematical expressions as abstract syntax trees (ASTs) and provides features based on symbolic analysis of those ASTs.
In this paper, we focus on two such features: \emph{\LaTeX{} output} and the \emph{constraint detection mechanism}.

JijModeling 2 is a complete rework that rebases the internal language on simply typed $\lambda$-calculus, mainly to support bound variables and the desugaring of comprehension syntax properly.
This higher-order representation makes the symbolic core more principled, but it also complicates the two features above: user-facing \LaTeX{} should recover conventional mathematical notation, and constraint detection should reason across parametrized families of constraints.

\begin{listing}[bp]
  \centering
  \begin{pycode}[linenos,autogobble,escapeinside=||]
  @jm.Problem.define("TSP, Decorated", sense=jm.ProblemSense.MINIMIZE) |\label{line:deco}|
  def tsp(problem: jm.DecoratedProblem):
      C = problem.CategoryLabel(description="Labels of Cities")
      N = C.count()
      x = problem.BinaryVar(dict_keys=(N, C),
          description="$x_{t,i} = 1$ if City $i$ is visited at time $t$")
      d = problem.Float(
          dict_keys=(C, C), description="distance between cities")
      problem += jm.sum(
          d[i, j] * x[t, i] * x[(t + 1) % N, j]
          for t in N for i in C for j in C) |\label{code:sum-compr}|

      problem += problem.Constraint("one time",
          [jm.sum(x[t, i] for t in N) == 1 for i in C]) |\label{code:constr-compr-1}|
      problem += problem.Constraint("one city",
          [jm.sum(x[t, i] for i in C) == 1 for t in N]) |\label{code:constr-compr-2}|
  \end{pycode}
  \caption{TSP, formulated in JijModeling\label{code:tsp-jm}}
\end{listing}

\Cref{code:tsp-jm} shows the TSP formulated in JijModeling 2.
This defines only a symbolic representation of the optimization problem; concrete input data is supplied later.
We call a variable that is later substituted with input data a \emph{placeholder}; in \Cref{fig:tsp}, $C$ and $d$ are examples.
In practice, such input data may have size 10,000 or more, and parametrized constraints like ``one city'' and ``one time'' are multiplied accordingly after substitution.
At the symbolic level, however, they remain single variables or constraints.
This lets us analyze mathematical expressions before expansion and reuse the same formulation for different input data.
In what follows, by ``the number of the parameters'' or ``the size of the problem'', we refer to the \emph{symbolic} size \emph{before substitution}.

The rest of the paper is organized around the following contributions.
\Cref{sec:syntax} gives an overview of the JijModeling 2 syntax and its egglog encoding, including the locally named representation of bound variables.
\Cref{sec:latex} presents an ensugaring procedure that reconstructs comprehension syntax from higher-order stream terms, followed by an egglog optimization pass that reduces temporary rebindings in the rendered \LaTeX{}.
\Cref{sec:detection} presents egglog-based constraint detection: first, a declarative replacement for the previous egg-based multi-stage detection pipeline~\cite{Ishii:aa}; second, a Henkin constant-like encoding of parametrized constraints that propagates side conditions; and third, a practical use of ensugaring to break large domain-set conditions into smaller rule premises before saturation.
\Cref{sec:concl} concludes.

\section{Syntax of JijModeling 2\label{sec:syntax}}

\begin{figure}[tbhp]
  \small
  \centering
    \nonterms{expr,pattern,primop}
  \caption{Syntax of JijModeling 2 (excerpt)\label{fig:syntax}}
\end{figure}

\Cref{fig:syntax} shows a part of the syntax of JijModeling 2, which is a simply typed $\lambda$-calculus extended with several constructs.
Built-in operations include (overloaded) arithmetic operations and operations on multidimensional arrays, dictionaries and sets.
In our codebase, bound variables are treated in the Locally Nameless Representation~\cite{Chargueraud:2012aa} with some user-supplied name hints on a $\lambda$-binder.

\subsection{Encoding in egglog\label{sec:egglog-encoding}}

\begin{listing}[htbp]
\begin{egglogcode}
(datatype Var (MkVar String i64 i64))
(datatype Expr (FVar String) (BVar Var) (Lam Pat Expr) ...)
(datatype Exprs (Nil) (Cons Expr Exprs))
\end{egglogcode}
\caption{Egglog-encoded AST (excerpt)\label{code:egglog-syntax}}
\end{listing}

\Cref{code:egglog-syntax} shows the egglog-encoded JijModeling expression AST (\eggloginline{Expr}).
It is almost a straightforward translation into egglog, except for \emph{$\lambda$-abstraction} and \emph{bound variables}.
As the proper treatment of bound variables, such as Slotted e-Graph~\cite{10.1145/3729326}, is not yet implemented in egglog, we adopt Locally \emph{Named} Representation~\cite{10.5555/645891.756642}, separating free and bound variables while naming both types of variables by names.
In particular, a free variable (\eggloginline{FVar}) is referred to solely by its (unique) name, whereas a bound variable (\eggloginline{Var} and \eggloginline{BVar}) is represented by a triple \eggloginline{(MkVar name offset nonce)}, where \eggloginline{name} is a textual name, \eggloginline{offset} is a de Bruijn-level like offset, and \eggloginline{nonce} is a random, unique identifier.
This representation also plays an important role in the constraint-detection in \Cref{sec:detection}.
Currently, our formulation in egglog does not support $\alpha$- or $\beta$-equivalence.
This is not a problem for our applications, as our purpose is not to optimize the evaluation of the lambda expressions.
Still, we support a form of limited $\alpha$-equivalence by using the conditional rules abstracted over the terms, which would suffice for constraint detection, as discussed in \Cref{sec:detection}.

\section{\LaTeX{} Output Involving Higher-Order Terms\label{sec:latex}}

JijModeling has long provided a feature to render the mathematical model as human-readable \LaTeX{} code --- indeed, \Cref{fig:tsp} is almost a verbatim copy of the \LaTeX{} output generated from \Cref{code:tsp-jm} by today's JijModeling 2.
However, pretty-printing into conventional mathematical notation becomes much harder in JijModeling 2 because the internal language includes higher-order constructs such as $\lambda$-abstraction.

\begin{figure}[tbp]
  \centering
    \centering
    \small
    \nonterms{compr,comprClauses,comprClause,forClause}
  \caption{Comprehension Syntax in JijModeling 2\label{fig:comprehension-syntax}}
\end{figure}

Why do we need $\lambda$-abstraction?
There are several reasons, but we focus on \emph{comprehension syntax} in this section.
Since JijModeling is primarily an EDSL in Python, its Python bindings support \emph{comprehension syntax} to improve user convenience.
\Cref{fig:comprehension-syntax} shows the concrete syntax of comprehension.
More precisely, the bindings parse the source code of a function decorated with \pyinline{@Problem.define} or similar decorators, as in \Cref{line:deco} of \Cref{code:tsp-jm}, and desugar expressions written in the comprehension syntax (e.g. \Cref{code:sum-compr,code:constr-compr-1,code:constr-compr-2}) into expressions built from stream operations such as \textsf{map}, \textsf{flat\_map}, and \textsf{filter}.
The desugaring rules are standard, so we omit them here.

For example, consider the following comprehension expression:
\medskip
\begin{center}
  \begin{pycode}
jm.sum(x[i] * t[i,j] for i in N if i % 2 == 0 for j in M)
\end{pycode}
\end{center}
\medskip
This desugars into the following expression:
\medskip
\begin{pycode}
jm.sum(
    N.filter(lambda i: i % 2 == 0)
     .flat_map(lambda i: M.map(lambda j: (i, j)))
     .map(lambda i, j: x[i] * t[i, j]))
\end{pycode}
\medskip

The problem is that, If such a desugared expression is emitted directly as \LaTeX{}, it is not so readable:
\begin{multline}
\sum \mathsf{map} \Big(\lambda (i, j).\; x_i t_{i, j},\\
      \mathsf{flat\_map}\big(\lambda i.\; \mathsf{map}\left(\lambda j.\; (i, j), M\right),\\
        \mathsf{filter}\left(\lambda i.\; i \bmod 2 = 0, N\right)\big)
    \Big)
    \tag{\ensuremath{\star}}
    \label{eqn:cryptic-math}
\end{multline}
For users, the conventional mathematical notation is preferable:
\begin{gather}
  \sum_{\substack{i = 0\\i \bmod 2 = 0}}^{N - 1}\sum_{j=0}^{M-1} x_i t_{i,j}
  \tag{\ensuremath{\star\star}}
  \label{eqn:readable-math}
\end{gather}
For this reason, JijModeling 2 converts expressions represented as compositions of $\mathsf{flat\_map}$, $\mathsf{map}$, and $\mathsf{filter}$ back into comprehension syntax; in other words, it performs a kind of ``\emph{ensugaring}''.
\begin{figure*}[tbh]
  \centering
  \small
  \drules[ES]{$\llbracket e \rrbracket = [e \ \mid \  \plural{\kappa}]$: ensugaring expressions into comprehensions}{Compr}{%
    Comp-Map,Comp-FlatMap,Comp-Filter
  }
  \drules[ES]{$\left\{\!\middle|\ p \in e\ \middle|\!\right\} = \plural{\kappa}$: ensugaring membership relations into clauses on the right-hand side of comprehensions}{Stmt}{%
    Stmts-Map,Stmts-FlatMap,Stmts-Filter,Stmts-Fallback
  }
  \caption{Ensugaring rules for comprehension syntax. Here, ${}^{\setminus \plural{a}} e$ denotes the expression obtained by substituting the free variables $\plural{a}$ for the outermost bound variables of $e$.\label{fig:comprehension-ensugar}}
\end{figure*}
\Cref{fig:comprehension-ensugar} shows a part of the ensugaring rules.
$\llbracket e \rrbracket$ denotes the result of ensugaring the expression $e$ into comprehension syntax, and $\left\{\!\middle|\ p \in e\ \middle|\!\right\}$ denotes the sequence of clauses on the right-hand side of a comprehension corresponding to the membership relation $p \in e$.
These rules temporarily assign all values bound by pattern matching to fresh variables and recursively reconstruct the right-hand side of the comprehension.
Since this process requires variables to be rebound, the transformation targets an extended comprehension syntax that introduces \textbf{let}-clauses, which were not permitted during desugaring.

However, this transformation alone introduces temporary variables at every stage, producing expressions that are cumbersome for users.
For example, ensugaring~\eqref{eqn:cryptic-math} again yields the following:
\[\small
\sum _{\substack{{i}_{2}=0\\{i}_{2}\bmod 2=0\\{i}_{1}={i}_{2}}}^{N-1}{\sum _{\substack{{j}_{1}=0\\\left\langle i,j\right\rangle =\left\langle {i}_{1},{j}_{1}\right\rangle }}^{M-1}{{x}_{i}  {t}_{i,j}}}
\]
To avoid this, JijModeling introduces a process that reduces the number of bound variables and \textsf{let}-clauses as much as possible.
This can be regarded as a program optimization pass, and use an extraction mechanism of egglog to achieve it: among equivalent comprehensions, choose one with fewer temporary variables and intermediate bindings while taking equivalence relations among bound variables into account.

\begin{listing}[tbp]
\begin{egglogcode}
(datatype Condition (If Expr) (Let Expr Expr) (Mem Expr Expr))
(datatype Conditions (Nil) (Cons Condition Conditions))
(datatype Comprehension (MkComprehension Expr Conditions))

(relation concludes (Expr))
(relation member (Expr Expr))
(relation assume (Condition))
(relation assumeMany (Conditions))

(with-ruleset ensugar-basic
  (rule ((assumeMany (Cons c cs))) ((assume c) (assumeMany cs)))
  (rule ((member (Tuple (Cons l ls)) (Tuple (Cons r rs))))
    ((member l r) (member (Tuple ls) (Tuple rs))))
  (rule ((assume (Mem l r))) ((member l r)))
  (rule ((assume (Let (Tuple (Cons p ps)) (Tuple (Cons e es)))))
    ((assume (Let p e))    (assume (Let (Tuple ps) (Tuple es)))))
  (rewrite ps es :when ((= (Tuple ps) (Tuple es))))
  (rewrite v  u  :when ((= (BVar v) (BVar u)))))

(with-ruleset comprehension
  (rule ((= t (MkVar v i n))) ((set (cost_table_MkVar v i n) i))))

(with-ruleset equate-and-dedupe
  (rewrite (BVar v) (BVar u) :when ((assume (Let (BVar v) (BVar u)))))
  (rewrite c cs :when ((= c (Cons (Let x x) cs)))))

(rule
  ( (concludes (BVar v)) (!= (BVar v) e)  (= let_var (Let (BVar v) e))
    (assumeMany (Cons let_var (Nil ))) )
  ( (union (BVar v) e)  (subsume (BVar v))
    (union (Cons let_var (Nil )) (Nil )) )
  :ruleset inline-trailing)

\end{egglogcode}
\caption{Optimization rules run after ensugaring\label{code:egglog-ensugar}}
\end{listing}

\Cref{code:egglog-ensugar} shows the core egglog rules for this process.
Concretely, the cost function is configured so that fewer \textbf{let}-bindings are preferred and outermost variables have lower cost; the minimum-cost term in the congruence closure is then extracted and used as the result of \LaTeX{} output.
The optimization pass processes each ensugared comprehension independently.
The relation \eggloginline{concludes} declares the left-hand-side term of the comprehension, whereas \eggloginline{Condition} corresponds to a single side condition listed on the right-hand side of the comprehension, and \eggloginline{Conditions} is a sequence of such conditions expressed as a linked-list.
While optimizing the comprehension, the facts in the RHS are recursively \eggloginline{assume}d, and each \eggloginline{Condition} is then registered into the fact base.

The optimization pass proceeds in four steps.
First, we saturate the e-graph with \eggloginline{basic-ensugar} and \eggloginline{comprehension} rules to register all facts involved in the comprehension.
Second, we equate variables and remove identical bindings by applying the \eggloginline{equate-and-dedupe} rules up to a fixed point.
Third, we saturate with the \eggloginline{inline-trailing} rule to inline trailing variables.
Finally, we extract the minimum-cost term from the e-graph as the result of \LaTeX{} output.
We use egglog's custom cost model support here, and variables with \emph{larger} offsets are given \emph{higher} cost, so the optimization pass prefers to keep variables with smaller offsets.
We assign smaller offsets to variables that are bound later in the comprehension conditions, making the last-bound variables tend to outlive earlier ones.
This is because variables bound earlier tend to be introduced by the intermediate closures generated in the middle of desugaring, and thus are more likely to be rebound later by the final \texttt{map} call.

With these rules, the optimization pass can optimize the ensugared term into the conventional form shown in~\eqref{eqn:readable-math}.
The resulting implementation runs almost instantly for small problems and takes \~5sec for practical problems with 50 variables and 30 constraints.

\paragraph{Other applications in \LaTeX{} output}
We are also using the database functionality of egglog to improve the readability of \LaTeX{} output in other ways.
For example, if the summation range is a natural number, we display it as $\sum_{i = 0}^{N - 1} x_i$ instead of $\sum_{i \in N} x_i$.
We also allow users to register preferred custom \LaTeX{} representations of free variables to improve brevity, and use that information when rendering.

\section{Declarative Constraint Detection with Side Condition Propagation\label{sec:detection}}

In this section, we discuss the application of egglog to \emph{constraint detection} in JijModeling.
As discussed in our previous work~\cite{Ishii:aa}, constraint detection problem is to detect certain forms of constraints in the mathematical model and invokes dedicated APIs provided by individual solvers, which can substantially speed up solving.
For example, an \emph{SOS1 constraint}, which appears in problems such as scheduling, is a constraint on a set of nonnegative variables $\set{x_1, \ldots, x_n}$ requiring that at most one of them be positive.

As an instance, consider the example below:
\begin{gather}
  \begin{gathered}
    (\text{even}) \quad \sum_{\substack{i < N\\i \bmod 2 = 0}} \delta_i \leq 1, \qquad
    (\text{odd}) \quad \sum_{\substack{i < N\\i \bmod 2 = 1}} \delta_i \leq 1 \\
    (\text{ub}) \quad x_i \leq M_i \delta_i \quad \forall i < N,
  \end{gathered}
  \tag{\ensuremath{\spadesuit}}
  \label{fig:separated-sos1}
\end{gather}
where $\delta_i \in \set{0, 1}$ and $x_i \in [0, \infty)$.
There, (even) and (odd) are two single constraints, whereas (ub) is a constraint parametrized over $i$ such that $i < N$.
In JijModeling 2, the constraint (ub) is represented as a pair of a domain set $\Set{ i \in \mathbb{N} | i < N}$ and a function from an index to the constraint instance $\lambda i.\; x_i \leq M_i \delta_i$.

In \eqref{fig:separated-sos1}, together with (ub), (even) and (odd) constitute SOS1 constraints on $\Set{x_i | i \bmod 2 = 0}$ and $\Set{x_i | i \bmod 2 = 1}$, respectively.
SOS1 constraints on general nonnegative variables are often declared in this way, as a combination of a single SOS1 constraint on binary variables (like (even) or (odd)) and a parametrized family of upper-bound constraints (like (ub)).

In the JijModeling 1 series, this detection functionality was implemented using egg~\cite{2021-egg}, the predecessor of egglog.
The intention was to use congruence closure by e-graphs to detect all algebraically equivalent patterns at once, instead of explicitly describing every pattern for every equivalent expression.
However, constraint detection with egg had difficulty handling constraints whose detection depends on detection results for multiple variable families, like the SOS1 example here.
In particular, it required a two-stage detection procedure: first detecting SOS1 constraints over binary variables with egg, taking those results out of egg, and then using that information to detect constraints over general nonnegative variables with egg again.

\begin{listing}[tbp]
\begin{egglogcode}
(rule ((<= (sum conds v[i]) 1) (binary_vars v)) ((sos1 conds v[i])))
(rule 
  ((sos1 conds b[i]) (binary_vars b) (<= v[i] (* M[i] b[i])) 
   (<= 0 v[i]) (<= v[i] M[i]))
  (sos1 conds v[i]))
\end{egglogcode}
\caption{Egglog rules for SOS1 constraint detection\label{code:egglog-sos1}}
\end{listing}
By contrast, the datalog-style rule-based inference mechanism in egglog lets us express constraint patterns such as SOS1 constraints entirely within egglog, as shown in \Cref{code:egglog-sos1}.

To see how the rules in \Cref{code:egglog-sos1} work, we outline constraint detection with egglog taking \eqref{fig:separated-sos1} as an example.

First, for each bound variable appearing in the problem, we generate a unique named bound variable from the pair consisting of its source-level name and a random number, as described in \Cref{sec:egglog-encoding}.
Then, for each such bound variable, we also register the facts assumed in their side conditions.
At the same time, the parametrized constraint is encoded as an egglog rule, conditioned on the domain set condition.
For example, \eqref{fig:separated-sos1} is converted into the following facts and registered in the egglog database:
\medskip
\begin{egglogcode}
; even
(< i_42 N)   (= (
(<= (sum ((< i_42 N) (= (
; odd
(< i_63 N)   (= (
(<= (sum ((< i_63 N) (= (
; ub
(rule ((< ?i N)) ((<= x[?i] (* M[?i] δ[?i]))))
(< i_98 N)
\end{egglogcode}
\medskip
In the above, \verb!i_42!, \verb!i_63!, and \verb!i_98! are the named bound variables with nonces, generated for the bound variable $i$ appearing in (even), (odd), and (ub), respectively.
In addition to the constraint itself, the facts $\texttt{i\_42} < N$ and $x_{\texttt{i\_42}} \mod 2 = k$, which correspond to the conditions on the summation index, are also registered in the egglog database.
For the parametrized (ub), egglog registers the conditional rule $i < N \implies c_i \leq M_i   \delta_i$; it also similarly generates the named bound variable \verb!i_98! and registers the fact $\texttt{i\_98} < N$.
From all these facts and derived rules, egglog now infers the following facts:
\medskip
\begin{egglogcode}
; from (even)/(odd) fact and sos1 rule
(sos1 ((< i_42 N) (= (
(sos1 ((< i_63 N) (= (

(<= x[i_42] (* M[i_42] δ[i_42])) ; from (even) and (ub)-rule
(<= x[i_63] (* M[i_63] δ[i_63])) ; from (odd) and (ub)-rule
\end{egglogcode}
\medskip
Then by the second rule in \Cref{code:egglog-sos1}, egglog can further detect the SOS1 constraints on even- and odd-sets of $x_i$s:
\medskip
\begin{egglogcode}
(sos1 ((< i_42 N) (= (
(sos1 ((< i_63 N) (= (
\end{egglogcode}
\medskip
In this way, unlike the existing detection procedure based on egg, even multi-stage constraint detection is completed entirely within egglog.

The constant \verb!i_42! in this example can be viewed as being introduced as something like a \emph{Henkin constant}\footnote{The concept of Henkin constants is used in proving the completeness theorem of classical first-order logic. This is closely related to the idea of Skolem constants, but plays a role of a \emph{conditional} witness for an existential formula, much like Hilbert's $\epsilon$-symbol.} for the formula $\varphi(i) \equiv i < N \land x_i \bmod 2 = 0$.
The difference from an ordinary Henkin constant is that the fact added to the database is not $\exists i \varphi(i) \implies \varphi(\texttt{i\_42})$, but the unconditional fact $\varphi(\texttt{i\_42})$.
This may appear contradictory when no such $i$ exists, in more complex cases.
However, in JijModeling, all bound variables are bounded, and these constants are used only as representatives for deriving schematic detections.
Before invoking a solver API, the detected pattern is interpreted back under the original bounded domain condition.
Thus, if the domain is empty, the representative facts do not produce concrete constraint instances, even though they exist in the egglog database.

\subsection{Further Optimization by Ensugaring}

For simple cases like \eqref{fig:separated-sos1}, the above procedure runs almost instantly.
However, we encountered a case where detection failed to complete within an hour.
After carefully analyzing it, we found that such a situation can occur when the constraint has more than a few parameters, and the domain set condition is expressed in terms of comprehensions.
One such example was the Deep Space Network Scheduling problem, as described in our tutorial documentation~\cite{jij-dsn}.
In particular, it contains the following constraint with somewhat gigantic side conditions:
\begin{multline*}
  {x}_{{n}_{1},m,{k}_{1},{t}_{1}}\cdot {x}_{{n}_{2},m,{k}_{2},{t}_{2}}=0\\
  \begin{split}
    &\forall ({n}_{1},{n}_{2},m,{k}_{1},{k}_{2},{t}_{1},{t}_{2})\;\text{s.t.}\\
    &\quad{n}_{1}\in \left\{0,\ldots ,N-1\right\},{n}_{2}\in \left\{0,\ldots ,N-1\right\},{n}_{2}\neq {n}_{1},\\
    &\quad m\in \left\{0,\ldots ,M-1\right\},\\
    &\quad {k}_{1}\in \left\{0,\ldots ,K-1\right\},{k}_{2}\in \left\{0,\ldots ,K-1\right\},\\
    &\quad {t}_{1}\in {AT}_{{n}_{1},m,{k}_{1}},{t}_{2}\in {AT}_{{n}_{2},m,{k}_{2}},\\
    &\quad ({t}_{1}-{su}_{{n}_{1}}\leq {t}_{2}-{su}_{{n}_{2}}\land {t}_{2}-{su}_{{n}_{2}}\leq {t}_{1}+{dr}_{{n}_{1}}+{td}_{{n}_{1}})\\
    &\quad (\lor {t}_{2}-{su}_{{n}_{2}}\leq {t}_{1}-{su}_{{n}_{1}}\land {t}_{1}-{su}_{{n}_{1}}\leq {t}_{2}+{dr}_{{n}_{2}}+{td}_{{n}_{2}})
  \end{split}
\end{multline*}

We simplified these side conditions into smaller ones for benchmarking, and it still took \~3 minutes to finish.
The confusing part was that egglog's runtime statistics reported that rule applications finished within \~100ms or so, contrary to the actual runtime.

{\sloppypar
It turned out that the root cause of this issue was that the egglog encoding of this domain set condition was simply far too huge.
In particular, the detection procedure described above encodes the above condition as a rule with \emph{just one single side condition}, \eggloginline{(member (Tuple n1 n2 m k1 k2 t1 t2) [HUGE HUGE TERM])}, where the domain set \eggloginline{[HUGE HUGE TERM]} is a deeply nested chain of \texttt{flat\_map}, \texttt{map}, and \texttt{filter} calls, given by the following:
\begin{multline*}
\mathsf{filter}\Big(
  \lambda (n_1,n_2,m,k_1,k_2,t_1,t_2).\;\\
  \quad
  (t_1 - su_{n_1} \leq t_2 - su_{n_2}
   \land t_2 - su_{n_2} \leq t_1 + dr_{n_1} + td_{n_1})\\
  \quad{}\lor{}
  (t_2 - su_{n_2} \leq t_1 - su_{n_1}
   \land t_1 - su_{n_1} \leq t_2 + dr_{n_2} + td_{n_2}),\\
  \mathsf{flat\_map}\Big(
      \lambda (n_1,n_2,m,k_1,k_2,t_1).\;\\
      \mathsf{map}\big(
        \lambda t_2.\; (n_1,n_2,m,k_1,k_2,t_1,t_2),
        AT_{n_2,m,k_2}
      \big),\\
    \mathsf{flat\_map}\Big(
      \lambda (n_1,n_2,m,k_1,k_2).\;
        \\\mathsf{map}\big(
          \lambda t_1.\; (n_1,n_2,m,k_1,k_2,t_1),
          AT_{n_1,m,k_1}
        \big),\\
      \mathsf{flat\_map}\Big(
        \lambda (n_1,n_2,m,k_1).\;
          \mathsf{map}\big(
            \lambda k_2.\; (n_1,n_2,m,k_1,k_2),
            K
          \big),\\
        \mathsf{flat\_map}\Big(
          \lambda (n_1,n_2,m).\;
            \mathsf{map}\big(
              \lambda k_1.\; (n_1,n_2,m,k_1),
              K
            \big),\\
          \mathsf{flat\_map}\Big(
            \lambda (n_1,n_2).\;
              \mathsf{map}\big(
                \lambda m.\; (n_1,n_2,m),
                M
              \big),\\
            \mathsf{filter}\Big(
              \lambda (n_1,n_2).\; n_2 \neq n_1,\\
              \mathsf{flat\_map}\big(
                \lambda n_1.\;
                  \mathsf{map}(
                    \lambda n_2.\; (n_1,n_2),
                    N
                  ),
                N
              \big)
            \Big)
          \Big)
        \Big)
      \Big)
    \Big)
  \Big)
\Big)
\end{multline*}
This observation and the trace log of egglog strongly suggested that egglog gets stuck processing this horrendously large rule \emph{before} the saturation or inference stage starts.
Hence, we applied the ensugaring procedure described in \Cref{sec:latex}, in particular $\left\{\!\middle|\ p \in e\ \middle|\!\right\}$, to \eggloginline{(member (Tuple n1 .. t2) [HUGE HUGE TERM])}.
This breaks up the side condition into pieces, and detection now successfully terminates within 3 seconds, which is a significant improvement.}

We also tried to apply the rebinding optimization on the ensugared side conditions, but it turned out that the optimization pass itself sometimes makes the runtime slightly worse, so we currently only apply the ensugaring transformation without the optimization pass.

The lesson is that the size of rule conditions matters before equality saturation begins: a rule can be too large for egglog to process efficiently even if the later saturation workload is small.

\section{Conclusion\label{sec:concl}}

We have discussed two applications of egglog in JijModeling 2: \LaTeX{} output for higher-order model terms and declarative constraint detection.
In both cases, egglog is useful not only for equational reasoning, but also as a database of derived facts over symbolic optimization models.
There are several future directions to explore, including the following:

\paragraph{Further runtime optimization in \LaTeX{} output} The current implementation takes a few seconds on practical problems; reducing this overhead remains future work.

\paragraph{Rigorous proof of soundness of Henkin-like detection approach} Although we are fairly confident that the Henkin-like constant trick in \Cref{sec:detection} is sound, we have not yet worked out a rigorous proof of it.
It would be good to have a formal proof of its soundness.

\paragraph{More thorough bound analysis in constraint detection} Propagating order-related side conditions is crucial for detection accuracy. Perhaps we can integrate interval arithmetic or affine arithmetic, but we have not yet figured out how to do it \emph{efficiently}.
  
\paragraph{Adoption in type-checking} We are planning to introduce a dependent type system in JijModeling 2, without full evaluation in the type-checking phase for security reasons.
We expect that egglog can play the role of a lightweight SMT-like engine to augment equality solving during dependent type checking.

\begin{acks}
  Many thanks to our colleagues at JIJ Inc, especially Yuya Nishimura, for the valuable discussions and for allowing me to submit this paper, even on a tight schedule.
  A part of this work was performed for Council for Science, Technology and Innovation (CSTI), Cross-ministerial Strategic Innovation Promotion Program (SIP), ``Promoting the application of advanced quantum technology platforms to social issues'' (Funding agency: QST).
\end{acks}

\printbibliography{}

\end{document}